\documentclass[preprint,showpacs,amsmath,amssymb]{revtex4}

\usepackage{graphicx}

\begin{document}
\title{Semi-Flexible Polymer in a Uniform Force Field in
Two Dimensions}
\author{
A. Lamura$^{1,2}$, T. W. Burkhardt$^{1,3}$ and G. Gompper$^{1}$ }
\affiliation{
$^1$ Institut f\"ur Festk\"orperforschung, Forschungszentrum
J\"ulich, 
D-52425 J\"ulich, Germany \\
$^2$ Istituto Nazionale per la Fisica della Materia
{\rm and} Dipartimento di Fisica, 
Universit\`a di Bari, {\rm and}
Istituto Nazionale di Fisica Nucleare, Sezione di Bari, 
via Amendola 173, 70126 Bari, Italy \\
$^3$ Department of Physics, Temple University, Philadelphia, PA
19122, USA }

\date{\today}

\begin{abstract}
The conformational properties of a semi-flexible polymer chain,
anchored at one end in a uniform force field, are studied in a
simple two-dimensional model. Recursion relations are derived for
the partition function and then iterated numerically. We calculate
the angular fluctuations of the polymer about the direction of the
force field and the average polymer configuration as functions of
the bending rigidity, chain length, chain orientation at the
anchoring point, and field strength.
\end{abstract}


\maketitle

\newpage

\section{Introduction}
\label{sec:intro}

The influence of external forces on the conformational properties
of polymers has been studied extensively in recent years. Polymers
stretched by attached magnetic beads \cite{smit92}, by laser
tweezers \cite{smit96,wang97}, and by optical fibers \cite{cluz96}
and polymers in flow fields \cite{perk95,lars97,smit99,lado00}
have received much of the attention \cite{bust00}. The study of
polymer deformation in elongational flow goes back to the
prediction of a coil-stretch transition \cite{genn74,genn79} and
early birefringence and light scattering experiments
\cite{full80,kell85}. Experimental techniques
\cite{perk95,perk94,kaes94,perk97} which allow direct
visualization of polymer conformations in simple flows have given
this field a new perspective. Here the main idea is to use
fluorescently labelled DNA molecules, which are long enough so
that their conformations can be resolved in an optical microscope.

In contrast to typical synthetic polymers, DNA chains are
semi-flexible, with a persistence length of about 80 nm. For
contour lengths of a few microns or more these chains behave as
flexible polymers, in the absence of external forces. In the case
of highly stretched DNA chains the bending rigidity has been shown
to play an important role \cite{bust94,mark95}.

The force-extension curve of a semi-flexible polymer pulled at
both ends is derived in Ref.~\cite{mark95}. Predicting the deformation
of a polymer in a flow field is considerably more complicated for
two reasons. First, there is a direct hydrodynamic interaction
between different polymer segments \cite{cifr99,rzeh99,rzeh00}.
Second, even if the conformation-dependent, fluctuating drag on
each bead is approximated by a friction term proportional to the
local flow velocity (``free-draining'' approximation), the force
on each bead depends on the positions of all other beads. Thus,
most theoretical studies have relied on computer simulations
\cite{lars97,cifr99,rzeh99,rzeh00,lars99} and/or consider flexible
chains \cite{cifr99,broc94,broc95,gg:gomp95f,wink97}.

In this paper we study the conformational properties of a
semi-flexible chain, anchored at one end, in two dimensions in a
constant force field. In our model the polymer partition function
is determined by simple recursion relations, which are easily
iterated numerically. Very little computing time is required, and
there is no statistical error in the results, but some other
approximations, such as the Villain approximation \cite{vill75},
are involved, as will become clear.

The two-dimensional model of a semi-flexible polymer is described
in Section~\ref{sec:model}. Recursion relations for the partition
function are derived in Section~\ref{sec:recursion}. In
Section~\ref{sec:fluctuations} we calculate the angular
fluctuations of the polymer segments about the direction of the
applied force, and in Section~\ref{sec:stretching} the longitudinal
extension due to the force. In Section~\ref{sec:theta0} we vary the
angle between the polymer and the force field at the anchoring point
and see how this affects the mean polymer configuration. Finally,
in Section~\ref{sec:end_pulled} the case of a polymer pulled at
its ends is briefly considered.

\section{The model}
\label{sec:model}

In the wormlike chain model of a semi-flexible polymer, the
Hamiltonian is given by \cite{doi86}
\begin{equation}
{\cal H}_0 = \frac{\kappa}{2} \int_0^L ds\;(\partial_s {\bf
t})^2\;,\label{H0}
\end{equation}
where ${\bf t}$ is a unit tangent vector and $s$ is the arc
length. We consider a discrete version of this model in two
spatial dimensions, with Hamiltonian
\begin{equation}
{\cal H}_0=J\sum_{i=1}^N(\theta_i-\theta_{i-1})^2\;.\label{ham}
\end{equation}
The polymer chain consists of $N+1$ line segments of fixed unit
length. The $i$th segment forms an angle $\theta_i$ with the $x$
axis. One end of the polymer is anchored at the origin, and the
orientation angle $\theta_0$ of the first segment is also assumed
to be fixed.

To include a uniform force field $F_0$ in the $x$ direction, we
add the terms
\begin{equation} \label{ham1}
{\cal H}_1=-F_0\sum_{i=1}^N x_i = -F_0\sum_{i=1}^N\sum_{j=1}^i
\cos \theta_j=-F_0\sum_{i=1}^N(N+1-i)\cos \theta_i
\end{equation}
to the Hamiltonian. The external field could be an electric,
gravitational, or uniform flow field. The partition function
corresponding to Eqs.~(\ref{ham}) and (\ref{ham1}) is given by
\begin{equation}
Z_N = \int_{-\infty}^\infty d\theta_1...\int_{-\infty}^\infty
d\theta_N \;
\exp\left\{-\sum_{i=1}^{N}\left[K(\theta_{i}-\theta_{i-1})^2
 + \hat{L}(N+1-i) \cos \theta_i \right]\right\}\;,
\label{part}
\end{equation}
where $K=J/kT$ and $\hat{L}=F_0/kT$.

A nice feature of this model is that it can be solved exactly in
the absence of an external field, i.e. for $\hat{L} = 0$. The mean
square end-to-end distance is given by
\begin{equation}
\langle R_N^2 \rangle = N+1 + \frac{2}{e^{1/\xi}-1} \left[N -
\frac{1-e^{-N/\xi}}{e^{1/\xi}-1} \right]
\end{equation}
with the persistence length $\xi = 4K$. For $N \gg \xi \gg 1$,
$\langle R_N^2 \rangle \simeq (1+2\xi)N$, corresponding to an
ideal flexible chain with Kuhn length $1+2\xi$. In the limit $N
\ll \xi$ with $\xi \gg 1$, $\langle R_N^2 \rangle \simeq (N+1)^2$,
corresponding to a rigid rod.

To obtain a more tractable model, we make the Villain
approximation \cite{vill75}
\begin{equation}
\exp\left[\hat{L}\cos \theta\right] \to \sum_{m=-\infty}^\infty
\exp\left[-L(\theta - 2 \pi m)^2\right]\;, \label{villain}
\end{equation}
in Eq.~(\ref{part}). It was originally introduced in studies of
the two-dimensional $x-y$ model and the roughening transition,
where the Hamiltonians have a similar form. The approximation
preserves the periodicity of the cosine function but leads to more
manageable Gaussian integrals. An irrelevant normalization factor
on the right-hand-side has been omitted. The constant $L$ may be
determined by expanding both sides of Eq.~(\ref{villain}) in
Fourier series and equating the lowest two Fourier coefficients.
This yields \cite{vill75}
\begin{equation}
L = \left[4 \ln \frac{I_0(\hat{L})}{I_1(\hat{L})}\right]^{-1}\;,
\end{equation}
where $I_0, I_1$ are modified Bessel functions.

Replacing $\sum_{m=-\infty}^\infty$ in Eq.~(\ref{villain}) by
$\sum_{m=-m_{\rm max}}^{m_{\rm max}}$ defines a further
approximation, which may be systematically improved by increasing
$m_{\rm max}$. In the finite $m_{\rm max}$ approximation,
configurations with up to $m_{\rm max}$ loops about the origin
receive the same statistical weight as for $m_{\rm max}=\infty$,
but the statistical weight of configurations with more than
$m_{\rm max}$ loops is underestimated.

\section{Recursion relations}
\label{sec:recursion}

As a first approximation we neglect all but the $m=0$ term in
Eq.~(\ref{villain}), replacing $e^{\hat{L}\cos\theta}$ by
$e^{-L\theta^2}$. This is a good approximation for sufficiently
large $K$ and/or $L$. The corresponding partition function is
\begin{eqnarray}
Z_N^0(h_1,&\dots&,h_N)=\int_{-\infty}^\infty
d\theta_1...\int_{-\infty}^\infty d\theta_N\nonumber\\
&\times&
\exp\left\{\sum_{i=1}^{N}\left[-K(\theta_{i}-\theta_{i-1})^2
 -L_i \theta_i^2+h_i\theta_i\right] \right\}\;,\label{part1}
\end{eqnarray}
where
\begin{equation} 
L_i=(N+1-i)L
\label{Li}\;,
\end{equation}
as follows from Eq.~(\ref{part}). The $h_i$ are auxiliary
variables that will be used in calculating thermal averages.

The partition function (\ref{part1}) may be evaluated by
straightforward integration over $\theta_1,\theta_2,\dots$. The
first $k-1$ integrations contribute
\begin{eqnarray}
Q_k^0(\theta_k)&=&\int_{-\infty}^\infty
d\theta_1...\int_{-\infty}^\infty d\theta_{k-1}\;
\exp\left\{\sum_{i=1}^{k}\left[-K(\theta_{i}-\theta_{i-1})^2
-L_i \theta_i^2+h_i\theta_i\right]\right\}\;,\nonumber\\
&=&q_k\exp\left[-\gamma_k\theta_k^2+2\beta_k\theta_k\right]\;,
\label{q2}
\end{eqnarray}
where $q_k$ is a constant, independent of $\theta_k$.
Equation~(\ref{q2}) and the recursive property
\begin{equation}
Q_k^0(\theta_k)=\int_{-\infty}^\infty d \theta_{k-1}\;\exp\left[-K
(\theta_k-\theta_{k-1})^2 -L_k\theta_k^2+h_k\theta_k\right]
Q_{k-1}^0(\theta_{k-1})\label{q3}
\end{equation}
imply
\begin{equation}
q_k=\left(\frac{\pi}{\gamma_{k-1}+K}\right)^{1/2}
    \exp\left[\frac{\beta_{k-1}^2}
        {\gamma_{k-1}+K}\right]q_{k-1}\label{q4}\;,
\end{equation}
where
\begin{eqnarray}
\gamma_1&=&K+L_1\;,\label{gam1}\\
\gamma_k&=&\frac{K\gamma_{k-1}}{\gamma_{k-1}+K}+L_k\;,\qquad
k=2,\dots,N\;,\label{gamk}
\end{eqnarray}
\begin{eqnarray}
\beta_1&=&K\theta_0+\frac{1}{2}h_1\;,\label{beta1}\\
\beta_k&=&\frac{K\beta_{k-1}}{\gamma_{k-1}+K}+\frac{1}{2}h_k\;,
        \qquad k=2,\dots,N\;.\label{betak}
\end{eqnarray}
Iterating Eq.~(\ref{q4}) to obtain $q_N$ and using $Z_N^0=\int
d\theta_NQ_N^0(\theta_N)=$\\
$=(\pi/\gamma_N)^{1/2}\exp\left[\beta_N^2/\gamma_N\right]q_N$, as
follows from Eqs.~(\ref{part1}) and (\ref{q2}), we obtain
\begin{eqnarray}
Z_N^0&=&\left[ \frac{\pi}{\gamma_N}
\;\frac{\pi}{\gamma_{N-1}+K}\dots \frac{\pi}{\gamma_1+K}\right
]^{1/2}\nonumber\\
&\times&\exp\left[\frac{\beta_N^2}{\gamma_N}+
\frac{\beta_{N-1}^2}{\gamma_{N-1}+K}+\dots+
\frac{\beta_1^2}{\gamma_1+K}\right]\;.\label{part2}
\end{eqnarray}

The partition function (\ref{part2}) is completely determined by
the recursion relations (\ref{gam1})-(\ref{betak}) and the $L_i$
defined in Eq.~(\ref{Li}).

\section{Angular fluctuations}
\label{sec:fluctuations}
We now derive the angular fluctuations
$\langle\theta_i^2\rangle$ of the polymer chain in a constant
force field from the partition function $Z_N^0$ of
Eqs.~(\ref{part1}) and (\ref{Li}). In this section we set the
initial angle $\theta_0$ and all of the auxiliary variables $h_i$
equal to zero. In this case
\begin{equation} \label{scalingform}
K\langle\theta_i^2\rangle=f_i(L/K, N)
\end{equation}
only depends on $K$ and $L$ in the combination $L/K$. This follows
from rescaling the angles $\theta_i$ in the partition function
(\ref{part1}). A second consequence is that the $\beta_k$ in
Eqs.~(\ref{beta1})-(\ref{part2}) all vanish.

Using the definition (\ref{Li}) of $L_i$, we write the recursion
relation (\ref{gamk}) for $\gamma_k$ in the form
\begin{equation}
\gamma_k=K+L(N+1-k)-\frac{K^2}{K+\gamma_{k-1}}\;.\label{gam3}
\end{equation}
According to Eqs.~(\ref{Li}), (\ref{part2}),  and (\ref{gam3}) the
angular fluctuations $\langle\theta_i^2\rangle=-\partial \ln
Z_N^0/\partial L_i$ satisfy
\begin{eqnarray}
\langle\theta_k^2\rangle &=&\frac{1}{2(\gamma_k+K)}+\left(
\frac{K}{\gamma_k+K}\right)^2 \langle\theta_{k+1}^2\rangle\;,
\quad k=1,\dots,N-1\;,
\label{recfl_0}\label{thetak}\\
\langle\theta_N^2\rangle &=&\frac{1}{2\gamma_N}\;.\label{thetaN}
\end{eqnarray}

\begin{figure}[ht]
\begin{center}
   \includegraphics*[width=0.7\textwidth]{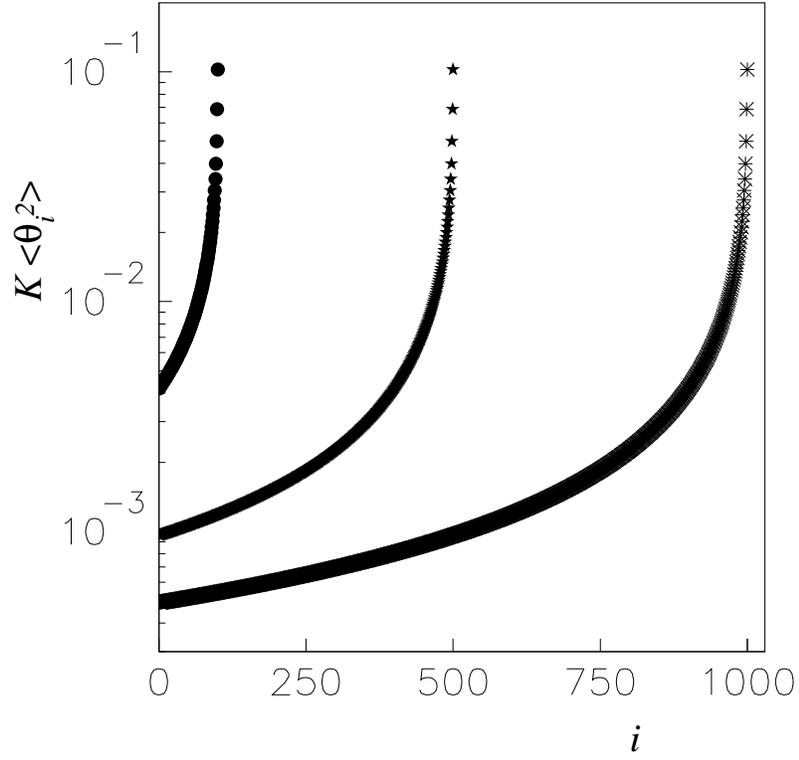}\\*
\end{center}
\caption{Average angular fluctuations
$K \langle\theta_i^2\rangle$ for $L/K=0.1$ and $N=100$
$(\bullet)$, $N=500$ $(\star)$, $N=1000$ $(\ast)$.} 
\label{fig1}
\end{figure}

We have calculated $\langle\theta_i^2\rangle$ by numerical
iteration of Eqs.~(\ref{gam1}), (\ref{gam3}), (\ref{thetak}), and
(\ref{thetaN}). The results for $L/K=0.1$ and $N=100$, $500$,
$1000$ are shown in Fig.~\ref{fig1}. For these three values of $N$
the results for $\langle\theta_N^2\rangle$ in Fig.~\ref{fig1} are
practically indistinguishable. The same is true of
$\langle\theta_{N-1}^2\rangle$, $\langle\theta_{N-2}^2\rangle$,
etc. The values $N=100$, $500$, and $1000$ are large enough so
that the angular fluctuations at the free end of the chain are
independent of the chain length.

\begin{figure}[ht]
\begin{center}
   \includegraphics*[width=0.7\textwidth]{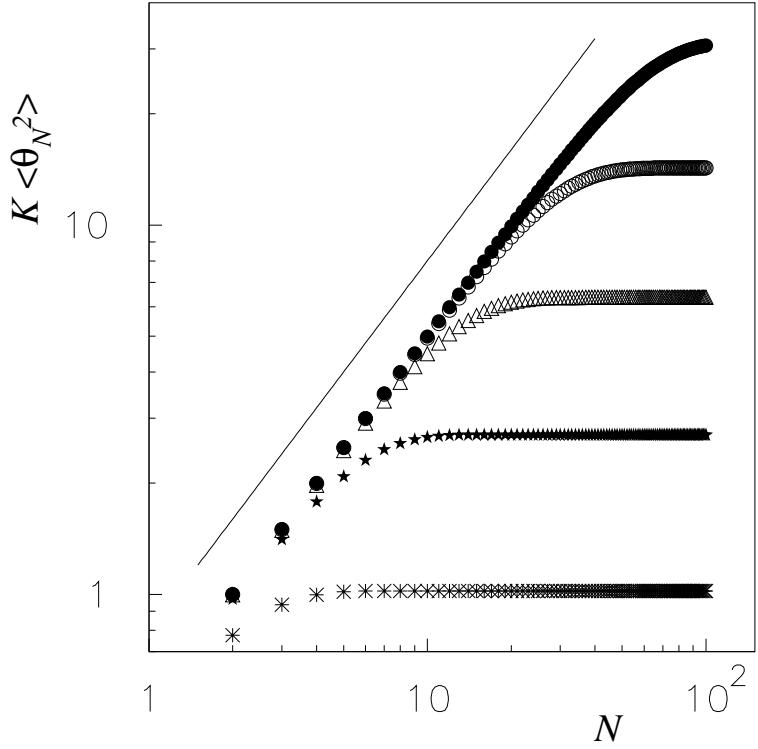}\\*
\end{center}
\caption{Average angular fluctuations
$K \langle\theta_N^2\rangle$ of the final chain segment as a
function of $N$ for $L/K=10^{-5}$ $(\bullet)$, $L/K=10^{-4}$
$(\circ)$, $L/K=10^{-3}$ $(\triangle)$, $L/K=10^{-2}$ $(\star)$,
$L/K=10^{-1}$ $(\ast)$. The straight line has slope 1.} 
\label{fig2}
\end{figure}

\begin{figure}[ht]
\begin{center}
   \includegraphics*[width=0.7\textwidth]{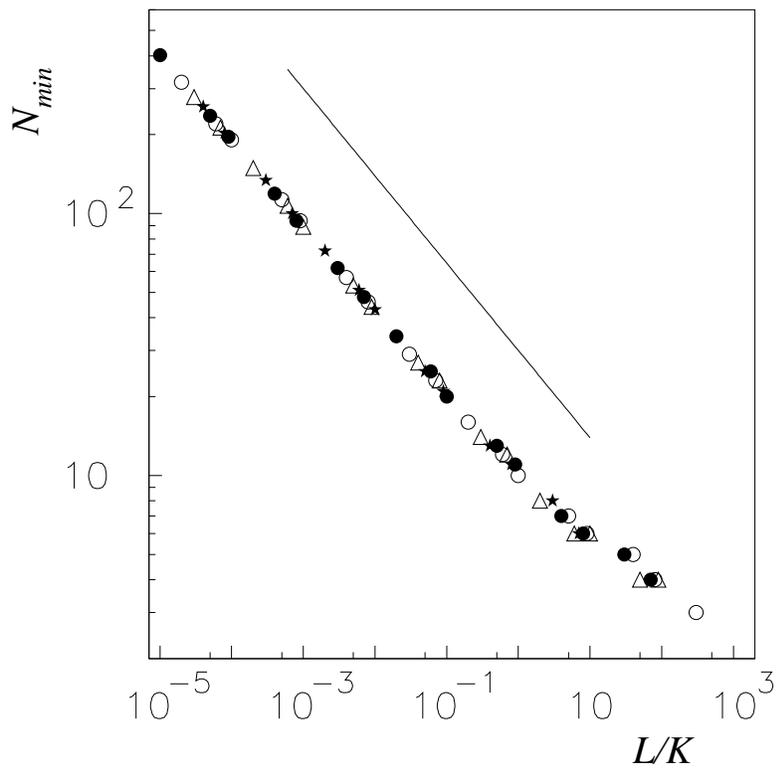}\\*
\end{center}
\caption{$N_{\rm min}$ as a function of $L/K$ for
$K=1$ $(\bullet)$, $K=10$ $(\circ)$, $K=100$ $(\triangle)$,
$K=1000$ $(\star)$. The straight line has slope -1/3.} 
\label{fig3}
\end{figure}

We now examine the $N$-dependence of the fluctuations. In
Fig.~\ref{fig2} the quantity $K \langle\theta_N^2\rangle$ is
plotted as a function of $N$ for five different values of $L/K$.
There is an obvious crossover from $N$-dependent to
$N$-independent behavior as $N$ increases. We denote the
approximate value of $N$ at the crossover by $N_{\rm min}$.
According to Eq.~(\ref{scalingform}), $N_{\rm min}$ only depends
on $K$ and $L$ in the combination $L/K$. Fig.~\ref{fig3} shows
$N_{\rm min}$ as a function of $L/K$. The data are in excellent
agreement with
\begin{equation}
N_{\rm min} \sim (L/K)^{-1/3},\quad L/K\ll 1\;.\label{Nmin1}
\end{equation}

The power law (\ref{Nmin1}) follows from the following argument:
For $L/K\ll 1$ the recursion relation (\ref{gam3}) implies
\begin{equation} \label{eq:gamma_a}
\gamma_k=\frac{K}{k}\left[ 1 + k A_k \frac{L}{K} +
O\left((L/K)^2\right)\right]
\end{equation}
where $A_k$ satisfies the recursion relation
\begin{equation} \label{eq:recursion_A}
A_k = \left( \frac{k-1}{k} \right)^2 A_{k-1} +
(N+1-k)
\end{equation}
with $A_1=N$. Since the factor multiplying $A_{k-1}$ in
Eq.~(\ref{eq:recursion_A}) approaches unity as $k$ increases, 
$A_N \sim N^2$ for $N \gg 1$. For fixed $L/K\ll 1$ and $N$ small, 
the first term
in $\gamma_N$, as given by Eq.~(\ref{eq:gamma_a}), is clearly the
dominant term, and $\gamma_N\simeq K/N$. However, the second term
becomes increasingly important as $N$ increases. It is reasonable
to assume that a crossover to a different $L/K$ dependence occurs
when the second term in $\gamma_N$, as given by Eq.~(\ref{eq:gamma_a}),
becomes comparable with the first, i.e. $NA_N\sim N^3\sim K/L$ for
$N\sim N_{\rm min}$. This leads to Eq.~(\ref{Nmin1}) and the
prediction  $\gamma_N\sim\gamma_{N_{\rm min}}\sim K(L/K)^{1/3}$
for $L/K\ll 1$, $N \gg N_{\rm min}$.

In the complementary regime $L/K \gg 1$ the recursion relation
(\ref{gam3}) implies
\begin{equation}
\gamma_N= K\left[\frac{L}{K}+1-\frac{K}{2L}
         +O\left((K/L)^2\right)\right]\label{gam4}
\end{equation}
for arbitrary $N$. Since this result is entirely independent of
$N$, the large $N$ behavior has its onset at
\begin{equation}
N_{\rm min}\sim 1,\quad L/K \gg 1\;.\label{Nmin2}
\end{equation}

We now derive exact analytic expressions for the fluctuations
$\langle\theta_{N}^2\rangle$, $\langle\theta_{N-1}^2\rangle,\dots$
at the end of an infinitely long chain. Reverse iteration of
Eq.~(\ref{gam3}), beginning with $\gamma_N$, leads to the
continued fraction
\begin{equation}
\gamma_N= K+L-\frac{K^2}{2K+2L\;-\ }\ \frac{K^2}{2K+3L\;-\
 }\ \frac{K^2}{2K+4L\;-\ }\ \dots\ \frac{K^2}{2K+NL}\;.
\label{confrac}
\end{equation}
In the limit $N\to\infty$ the continued fraction may be evaluated
with the help of Eq.~(9.1.73) in Ref.~\cite{abra70}, yielding
\begin{equation} \label{gaminfty}
\gamma_\infty=K\left[\frac{J_\nu(\nu)}
    {J_{\nu}^\prime(\nu)}-1\right]^{-1}\;,\quad \nu=\frac{2K}{L}\;.
\end{equation}
Here $J_\mu(z)$ is the standard Bessel function, and
$J_\mu^\prime(z)$ is its derivative with respect to $z$. From
Eq.~(9.3.23) in Ref.~\cite{abra70} and Eq.~(\ref{confrac}), we
obtain
\begin{equation}
\gamma_\infty\simeq
K\left\{\begin{array}{l}
\displaystyle{3^{1/3}
\frac{\Gamma(2/3)}{\Gamma(1/3)}
          \left(\frac{L}{K}\right)^{1/3}\;,}\quad\\
\\\displaystyle{\frac{L}{K}+1-\frac{K}{2L},}
              \quad\end{array}\right.\begin{array}{l}
\displaystyle{\frac{L}{K}\ll 1}\;,\label{smallL}\\ \\
\displaystyle{\frac{L}{K}\gg 1}\;.
\end{array}
\end{equation}
These limiting forms are consistent with the expressions for
$\gamma_\infty\sim\gamma_{N_{\rm min}}$ for small and large $L/K$
given in Eq.~(\ref{gam4}) and the paragraph which preceeds it.

>From the result (\ref{gaminfty}) for $\langle\theta_N^2\rangle=
\left(2\gamma_N\right)^{-1}$ in the large-$N$ limit, it is
straightforward to calculate $\langle\theta_{N-1}^2\rangle$,
$\langle\theta_{N-2}^2\rangle,\dots$ using
Eqs.~(\ref{gam3})-(\ref{thetaN}).

\begin{figure}[ht]
\begin{center}
   \includegraphics*[width=0.7\textwidth]{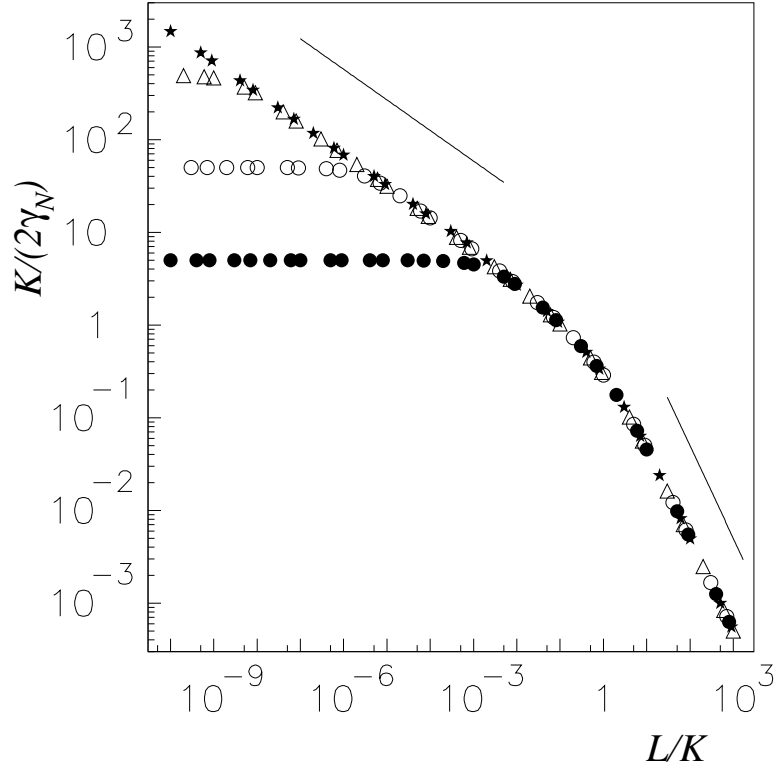}\\*
\end{center}
\caption{$K \langle\theta_N^2\rangle=K/(2\gamma_N)$ for a
polymer in a constant force field
as a function of $L/K$ for $N=10$ ($\bullet$), $N=100$ ($\circ$),
and $N=1000$ ($\triangle$), together with the exact result
(\ref{gaminfty}) for $N\to\infty$ ($\star$). The straight lines
have slopes $-1/3$ and $-1$, respectively.} 
\label{fig4}
\end{figure}

In Fig.~\ref{fig4}, $K\langle\theta_N^2\rangle=K/(2\gamma_N)$ is
plotted as a function of $L/K$ for $N=10$, $100$, $1000$, and
compared with the analytic prediction (\ref{gaminfty}) for
$N\to\infty$.

As stated in Section~\ref{sec:recursion}, the $m_{\rm max}=0$
approximation is accurate for sufficiently large $K$ and/or $L$.
One can use the results of this section to determine the domain of
validity more precisely. The approximation, i.e. replacing
$\cos\theta$ by $1-\frac{1}{2}\theta^2$, should be quite reliable
if, say,
$\langle\theta_N^2\rangle=\left(2\gamma_N\right)^{-1}<(\pi/4)^2$
or $\gamma_N>8/\pi^2\simeq 1$. Computing $\gamma_N$ by numerical
iteration, one can readily check whether this inequality is
satisfied for particular values of $K$, $L$, and $N$. According to
Eqs.~(\ref{eq:gamma_a}), (\ref{smallL}), and (\ref{gam4}), the
inequality $\gamma_N>1$ corresponds to $K>N$ for $L/K\ll 1$ and
$N\ll N_{\rm min}$, to $K(L/K)^{1/3}>1.4$ for $L/K\ll 1$ and $N\gg
N_{\rm min}$, and to $L>1$ for $L/K\gg 1$ and arbitrary $N$.

\section{Longitudinal stretching}
\label{sec:stretching}

For a polymer in a constant force field, one of the main
quantities of interest is the average extension in the flow
direction. According to a prediction of Marko and Siggia
\cite{mark95},
\begin{equation} \label{eq:x_extension}
\frac{\langle x_N \rangle}{N}  \simeq 1-C_0(LKN)^{-1/2}\;,
\end{equation}
with $C_0=1$. This result was derived by approximating the
restoring force at position $s$ along the chain with the
thermodynamic result for a polymer of length $s$ pulled at its
ends. Note, however, that a polymer in a flow field fluctuates
most strongly at the free end and not at all at the anchored end,
quite unlike a polymer pulled at its ends. The derivation also
assumes a boundary condition $ \langle \cos \theta_N \rangle \sim
LK^2 \ll 1$ at the end of the chain, at odds with the exact result
$\langle \theta_N^2 \rangle \sim (LK^2)^{-1/3}$ in Eqs.~(\ref{thetaN}) 
and (\ref{smallL}) for large $K$ and $N$, where our
discrete model is equivalent to the continuum model of
Ref.~\cite{mark95}. One advantage of our approach is that it avoids
these assumptions and yields numerically exact results for the
model with partition function (\ref{part1}).

We have checked Eq.~(\ref{eq:x_extension}) for our model, using
the relations
\begin{eqnarray}
\langle x_N \rangle &=&\sum_{j=0}^N\langle\cos\theta_j\rangle=
\frac{1}{2}\sum_{j=0}^N\langle
e^{i\theta_j}+e^{-i\theta_j}\rangle\;,\label{xN}\\
\langle e^{\pm i\theta_j}\rangle &=&\frac{Z_N^0(0,\dots,h_j=\pm
i,0,\dots)}{Z_N^0(0,\dots,0)}\;.\label{eitheta}
\end{eqnarray}
Here the numerator on the right side of Eq.~(\ref{eitheta}) is the
partition function of Eqs.~(\ref{part1}) and (\ref{part2}) with
all of the auxiliary fields set equal to zero except that $h_j=i$.
In the denominator all of the auxiliary fields, including $h_j$,
vanish.

Calculating these partition functions recursively using
Eqs.~(\ref{gam1}), (\ref{gamk}), and (\ref{part2}), with the $L_i$
defined by Eq.~(\ref{Li}), we obtained the results for the
extension shown in Fig.~\ref{fig5}. The data for sufficiently
large $K$ do indeed confirm Eq.~(\ref{eq:x_extension}). For our
model $C_0\simeq 0.23$.

It is instructive to compare the chain length $N$ with $N_{min}$
in the regime where Eq.~(\ref{eq:x_extension}) applies. For
$K=1000$ and $N=10\thinspace 000$ (filled circular points in
Fig.~\ref{fig5}) the six-decade interval $10^{0}<LKN<10^{6}$
corresponds to $10^{-10}<L/K<10^{-4}$, or (see Eq.~(\ref{Nmin1})
and Fig.~\ref{fig3}) to $20\thinspace 000>N_{min}>200$. Thus, the
inequality $N \gg N_{min}$ only holds for the last three decades
of $LKN$.

There are deviations from the straight line in Fig.~\ref{fig5} for
smaller $K$ and larger $L$. For $L \gg K$, the recursion relation
(\ref{gam3}) implies
\begin{equation}
\gamma_k\simeq (N+1-k)L\;,
\end{equation}
and
\begin{equation}
\langle x_N \rangle = \sum_{i=0}^N \langle \cos \theta_i \rangle
         \simeq N - \frac{1}{2} \sum_{i=0}^N \langle \theta_i^2 \rangle
         \simeq N - \frac{1}{4} \sum_{i=0}^N \frac{1}{\gamma_i}\;.
\end{equation}
The sum over the chain segments is easily carried out and yields
\begin{equation}
\langle x_N \rangle\simeq N-\frac{\ln N}{4L}.
\end{equation}

In Fig.~\ref{fig6} numerical data for several bending rigidities,
chain lengths and force fields are compared with this result. The
agreement is excellent. Thus, for a sufficiently strong force
field or a sufficiently small bending rigidity, the chain
extension varies as $L^{-1}$, just as for flexible chains with
arbitrary $L$.

\section{Mean polymer configuration for $\theta_0\neq 0$}
\label{sec:theta0}

In this section we consider polymer conformations with the first
segment fixed at a non-zero angle $\theta_0$ with respect to the
direction of the force field. We make the Villain approximation
and restrict the sum in Eq.~(\ref{villain}) to $m=-1,0,1$. In the
corresponding polymer partition function there is a separate sum
over $m$ for each polymer segment. For a sufficiently stiff
polymer and/or a sufficiently strong force field, the
conformations of the chain are dominated by the {\em same}
potential minimum. In this case the separate sums may be replaced
by a single sum, leading to the simpler partition function
\begin{eqnarray}
\tilde{Z}_N(\tilde{h}_1,&\dots&,\tilde{h}_N) =
\sum_{m=-1}^1\int_{-\infty}^\infty
d\theta_1...\int_{-\infty}^\infty d\theta_N \nonumber\\&\times&
\exp\left\{\sum_{i=1}^{N} \left[-K(\theta_{i}-\theta_{i-1})^2 -
L_i (\theta_i-2\pi m)^2
+\tilde{h}_i\theta_i\right]\right\}\;.\label{part3}
\end{eqnarray}
Here we have again introduced a set of auxiliary variables
$\tilde{h}_i$, to be used in constructing thermal averages.

Expanding $(\theta_i-2\pi m)^2$ in powers of $\theta_i$, one sees
that the partition function (\ref{part3}) can be expressed as
\begin{equation}
\tilde{Z}_N(\tilde{h}_1,\dots,\tilde{h}_N)=\sum_{m=-1}^1
e^{-4\pi^2m^2(L_1+\dots+L_N)}Z_N^0(\tilde{h}_1+4\pi
mL_1,\dots,\tilde{h}_N+4\pi mL_N)\label{part4}
\end{equation}
in terms of the $m=0$ partition function $Z_N^0(h_1,\dots,h_N)$
defined in Eq.~(\ref{part2}). Since we know how to calculate
$Z_N^0$ numerically with the recursion relations of 
Section~\ref{sec:recursion}, we
can also calculate $\tilde{Z}_N$ via Eq.~(\ref{part4}). Using
Eqs.~(\ref{Li}), (\ref{gam1})-(\ref{part2}), (\ref{part4}), and
the relations
\begin{eqnarray}
\langle x_k\rangle &=&\sum_{j=0}^k\langle\cos
\theta_j\rangle=\frac{1}{2}\sum_{j=0}^k\langle 
          e^{i\theta_j}+e^{-i\theta_j}\rangle\;,\\
\langle y_k\rangle &=&\sum_{j=0}^k\langle\sin
\theta_j\rangle=\frac{1}{2i}\sum_{j=0}^k\langle 
            e^{i\theta_j}-e^{-i\theta_j}\rangle\;,\\
\langle e^{\pm i\theta_j}\rangle
&=&\frac{\tilde{Z}_N(0,\dots,\tilde{h}_j=\pm i,0,\dots)}
             {\tilde{Z}_N(0,\dots,0)}\;,
\end{eqnarray}
we have evaluated the average position of the $j$th segment of the
polymer chain for fixed $\theta_0$.

Figs.~\ref{fig7}-\ref{fig12} show how the average position depends
on the tilt angle $\theta_0$, field strength $L$, and bending
rigidity $K$. In all of these figures $N=100$.

Figs.~\ref{fig7} and \ref{fig8} show the average polymer
configuration in the $x,y$ plane. As the field strength increases,
the polymer is bent towards the field direction and is stretched
longitudinally. The elongation is more pronounced for smaller
bending rigidities.

In Fig.~\ref{fig9} the transverse extension of the polymer as a
function of the tilt angle is shown in more detail. The curves are
sinusoidal for small field strengths $L$ but for larger $L$ bend
abruptly near $\theta_0=\pm\pi$, due to the instability of a
polymer directed against the force field. Since our model includes
fluctuations, the polymer "tunnels" between the two equivalent
free-energy minima, and there is no spontaneous symmetry breaking
at $\theta_0=\pi$.

Fig.~\ref{fig10} shows the tranverse extension as a function of
the field strength for three different $\theta_0$. For $K=10$,
$N=100$, and $L < 10^{-5}$, the effect of the force field is
negligible. For stronger force fields, there seems to be a regime
where $\langle y_N \rangle \sim L^{-2/5}$.

The contour length $C_l$ of the average configuration (see
Figs.~\ref{fig7} and \ref{fig8}) is shown in Fig.~\ref{fig11}.
Again, the effect of the force field is negligible for $K=10$,
$N=100$, and $L < 10^{-5}$ . Varying the tilt angle only affects
the contour length near the onset of the deformation due to the
force field.

Finally, we have considered the angle $\delta$ between a line
through the end-points of the chain and the direction of the force
field. A weak force field deforms the polymer only slightly, and
$\delta$ varies linearly with the tilt angle $\theta_0$. For a
strong force field, on the other hand, $\delta$ varies abruptly as
$\theta_0$ approaches $\pi$, due to the instability mentioned
above. The behavior of $\delta$ as a function of $L$ in
Fig.~\ref{fig12} is qualitatively similar to that of $\langle y_N
\rangle$ in Fig.~\ref{fig10}.

\section{Polymer pulled at its ends}
\label{sec:end_pulled}

Thus far we have considered an external force field that acts on
each monomer of the semi-flexible polymer. With only minor
modifications the case of a constant force applied at the ends of
the polymer can also be studied. Anchoring one end at the origin
at a fixed angle $\theta_0$, we apply a constant force at the
other end by replacing Eq.~(\ref{ham1}) with ${\cal
H}_1=-F_0x_N=-F_0\sum_{j=1}^N\cos\theta_j$. In the Villain
approximation (\ref{villain}) the partition function is again
given by Eq.~(\ref{part1}), but Eq.~(\ref{Li}) is replaced by
$L_i=L$. With this definition of the $L_i$ the partition function
may be calculated recursively, as in Section~\ref{sec:recursion}. For
symmetric boundary conditions at the ends of the chain, i.e.
$\theta_0$ and $\theta_N$ both fixed or both free to fluctuate,
the calculation is also straightforward.

For fixed $\theta_0=0$ and fluctuating $\theta_N$, the angular
fluctuations of the polymer segments are given by 
Eqs.~(\ref{scalingform}),
(\ref{thetak}), and (\ref{thetaN}), with Eq.~(\ref{gam3}) replaced by
\begin{equation}
\gamma_k=K+L-\frac{K^2}{K+\gamma_{k-1}}\;.
\label{gamends}
\end{equation}
In the long chain limit $\gamma_N$ approaches the fixed point
\begin{equation}
\gamma_\infty=\frac{L}{2}
             +\left[\left(\frac{L}{2}\right)^2+KL\right]^{1/2}
\label{fixpoint}
\end{equation}
of Eq.~(\ref{gamends}).

For large $K$ and $N$ our discrete model is equivalent to the
continuum model of  Marko and Siggia \cite{mark95}. In this limit
Eqs.~(\ref{thetaN}) and (\ref{fixpoint}) imply the same result
$\langle\theta_N^2\rangle=(4KL)^{-1/2}$ for the angular
fluctuations as in Ref.~\cite{mark95}.

In Fig.~\ref{fig13}, $K\langle\theta_N^2\rangle=K/(2\gamma_N)$ is
plotted as a function of $L/K$ for $N=10$, $100$, $1000$, and
$10\thinspace 000$ and compared with the analytic prediction
(\ref{fixpoint}) for $N\to\infty$.

We have also calculated the mean configuration of a polymer pulled
at its ends for fixed $\theta_0 > 0$ and fluctuating $\theta_N$.
Fig.~\ref{fig14} shows the tranverse extension as a function of
the force for three different $\theta_0$. For $K=10$, $N=100$, and
$L < 10^{-3}$, the effect of the force is negligible. For stronger
forces there seems to be a regime where $\langle y_N \rangle \sim
L^{-2/5}$. We found quite similar behavior for a polymer in a
uniform force field, as shown in Fig.~\ref{fig10}.

\section{Concluding remarks}
\label{sec:conclusions}

For calculating the conformational properties of a semi-flexible
chain in a uniform force field our recursive approach has several
advantages: (i) It requires very little computing time, and (ii)
it allows one to consider very long chains. For a clearly-defined
model exact numerical results are obtained. Thus, (iii) there is
no statistical error, and (iv) some of the approximations in
earlier theoretical work are avoided. Finally, (v) the recursion
relations furnish some analytical insight. We were able to obtain
some exact results for the asymptotic behavior of long chains.
While most previous studies have focused on the force-extension
relation, we have also analyzed angular fluctuations.

A disadvantage of the approach is the limitation to two
dimensions. However, many of the results probably apply, at least
qualitatively, to chains in three spatial dimensions. Furthermore,
the results are directly applicable to polymers confined to two
dimensions, for example, DNA electrostatically bound to fluid
lipid membranes \cite{maie99}.

The Villain approximation was used to obtain a tractable model. It
preserves the periodicity in $\theta$ and is no more unrealistic
than using a quadratic bending energy for arbitrary angles or
ignoring excluded volume. We only presented results for the
single-$m$ approximation with $m_{\rm max}\leq 1$, which
underestimates the statistical weight of configurations with
segments pointing in widely different directions but is accurate
for sufficiently large $K$ and/or $L$.

The single-$m$ approximation can, of course, be improved at the
cost of greater computing time. Retaining all the Villain sums
leads to the partition function
\begin{eqnarray}
{\cal Z}_N(\tilde{h}_1,\dots,\tilde{h}_N)&=&
\sum_{m_1=-\infty}^\infty\dots \sum_{m_N=-\infty}^\infty
e^{-4\pi^2m_1^2L_1}\dots e^{-4\pi^2m_N^2L_N}\nonumber\\
&&\times\ Z_N^0(\tilde{h}_1+4\pi m_1L_1,\dots,\tilde{h}_N+4\pi m_N
L_N)\label{part5}
\end{eqnarray}
instead of Eq.~(\ref{part4}). Here $Z_N^0$ is the $m=0$ partition
function in Eq.~(\ref{part2}), which we know how to compute
recursively. Usually one is interested in $K$ and $L$ for which
the angular differences between adjacent segments are small,
certainly less than $2\pi$. Then the sums on the right side of
Eq.~(\ref{part5}) may be restricted to the terms with $-1\leq
m_1\leq 1$, $m_1-1\leq m_2\leq m_1+1$, etc. Computing ${\cal Z}_N$
with no further approximations requires $3^N$ evaluations of
$Z_N^0$.

\section{Acknowledgements}
Helpful discussions with R. Winkler and U. Seifert are gratefully
acknowledged. T.W.B. thanks the Institut f\"ur
Festk\"orperforschung, Forschungszentrum J\"ulich for hospitality
and the Alexander von Humboldt Stiftung for financial support.

\bibliographystyle{prsty}
\bibliography{amphiphile,gompper}

\newpage

\begin{figure}[ht]
\begin{center}
   \includegraphics*[width=0.7\textwidth]{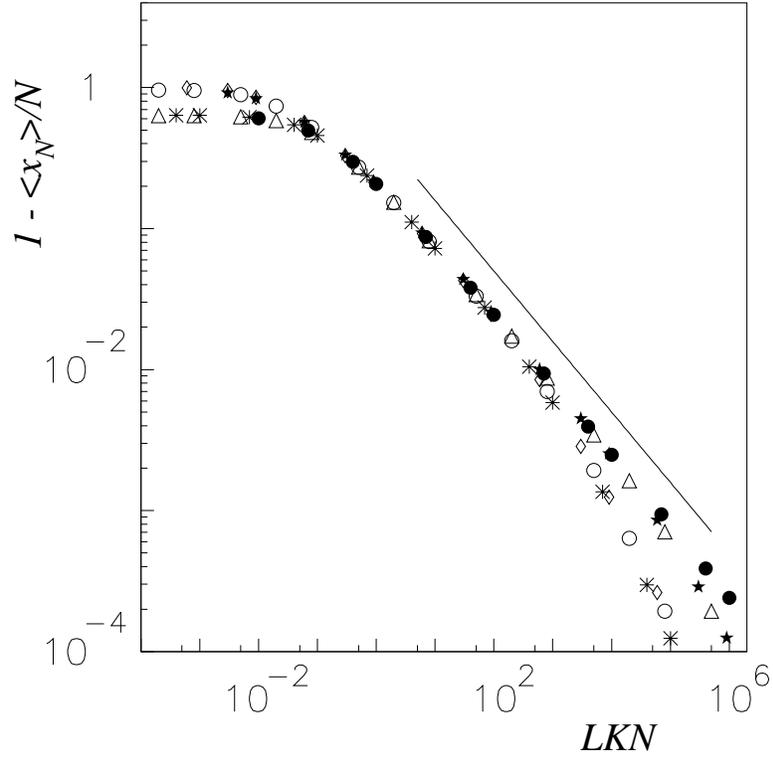}\\*
\end{center}
\caption{Average extension,
$\langle x_N \rangle$ of the chain in the direction of the force
field direction as a function of the field strength $L$, for
$K=10$, $N=100$ ($\ast$); $K=10$, $N=1000$ ($\circ$); $K=10$,
$N=10000$ ($\diamond$); $K=100$, $N=1000$ ($\triangle$); $K=100$,
$N=10000$ ($\star$); $K=1000$, $N=10\thinspace 000$ ($\bullet$).
The straight line has slope $-1/2$.} 
\label{fig5}
\end{figure}

\begin{figure}[ht]
\begin{center}
   \includegraphics*[width=0.7\textwidth]{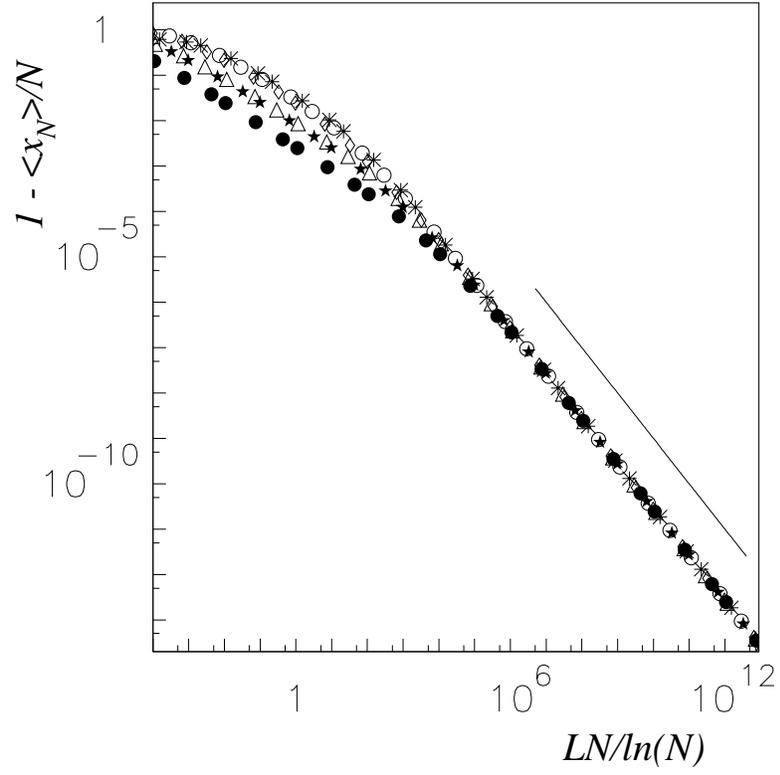}\\*
\end{center}
\caption{Average extension $\langle x_N \rangle$ of
the chain in the direction of the force field as a function of the
field strength $L$. The symbols correspond to the same parameters
as in Fig.~\ref{fig5}. The straight line has slope $-1$.} 
\label{fig6}
\end{figure}

\begin{figure}[ht]
\begin{center}
   \includegraphics*[width=0.7\textwidth]{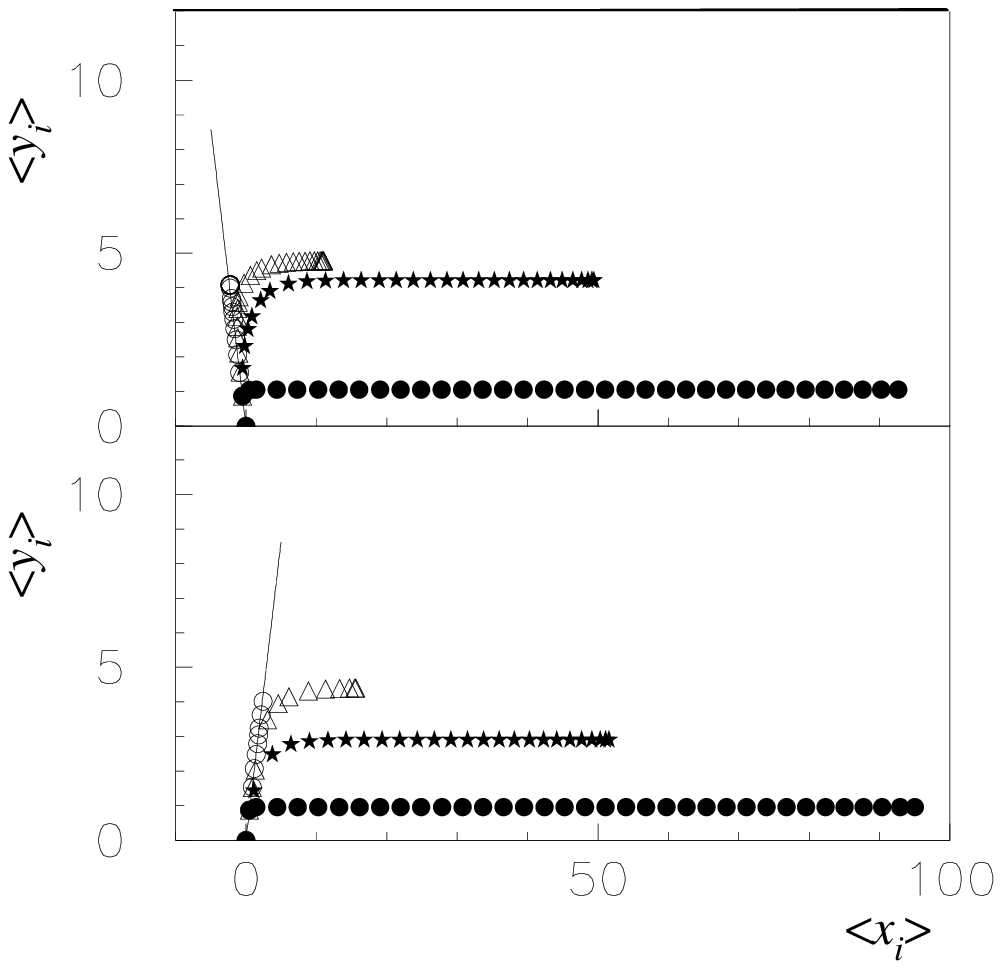}\\*
\end{center}
\caption{Average positions
$\langle x_i\rangle,\langle y_i\rangle$ of a chain of length
$N=100$ with $K=1$, $\theta_0=\pi/3$ (lower panel), and
$\theta_0=2 \pi/3$ (upper panel), with $L=10^{-5}$ $(\circ)$,
$L=10^{-4}$ $(\triangle)$, $L=10^{-3}$ $(\star)$, $L=10^{-1}$
$(\bullet)$. The straight line indicates the tilt angle
$\theta_0$.} 
\label{fig7}
\end{figure}

\begin{figure}[ht]
\begin{center}
   \includegraphics*[width=0.7\textwidth]{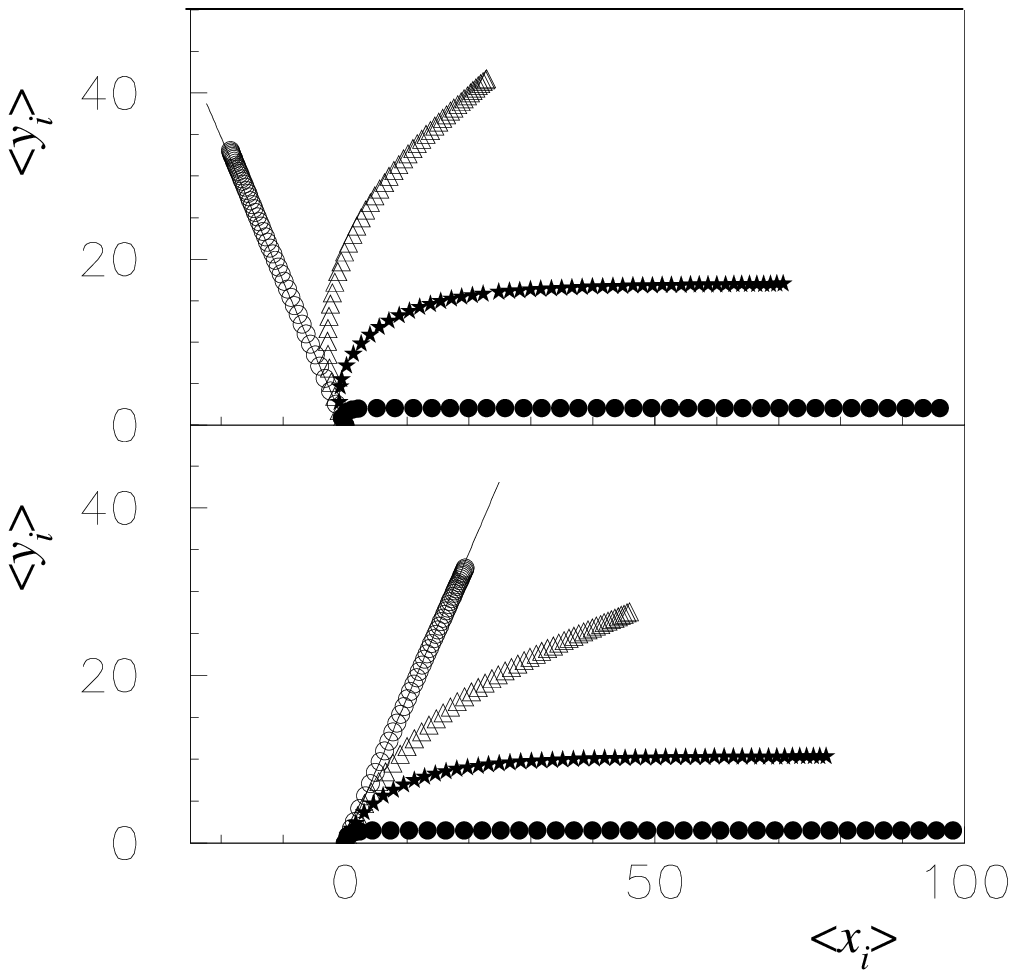}\\*
\end{center}
\caption{Average positions
$\langle x_i\rangle,\langle y_i\rangle$ of a chain of length
$N=100$ with $K=10$, $\theta_0=\pi/3$ (lower panel), and
$\theta_0=2 \pi/3$ (upper panel), with $L=10^{-5}$ $(\circ)$,
$L=10^{-4}$ $(\triangle)$, $L=10^{-3}$ $(\star)$, $L=10^{-1}$
$(\bullet)$. The straight line indicates the tilt angle
$\theta_0$.} 
\label{fig8}
\end{figure}

\begin{figure}[ht]
\begin{center}
   \includegraphics*[width=0.7\textwidth]{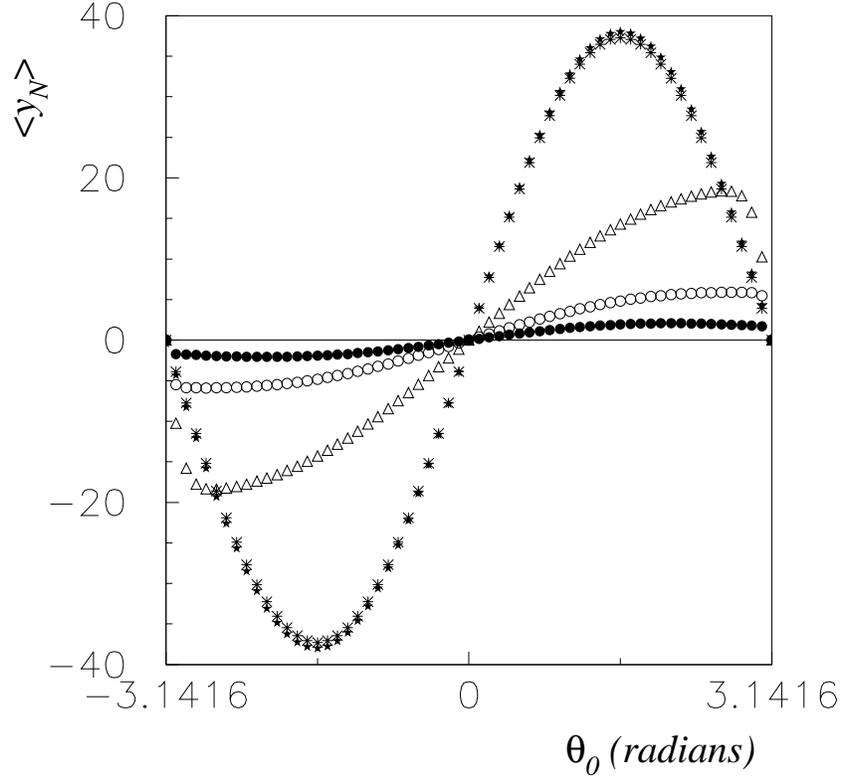}\\*
\end{center}
\caption{Average transverse extension
$\langle y_N\rangle$ as a function of $\theta_0$ for $N=100$,
$K=10$ and $L=10^{-9}$ $(\ast)$, $L=10^{-5}$ $(\star)$ , and
$L=10^{-3}$ $(\triangle)$ , $L=10^{-2}$ $(\circ)$ , $L=10^{-1}$
$(\bullet)$.} 
\label{fig9}
\end{figure}

\begin{figure}[ht]
\begin{center}
   \includegraphics*[width=0.7\textwidth]{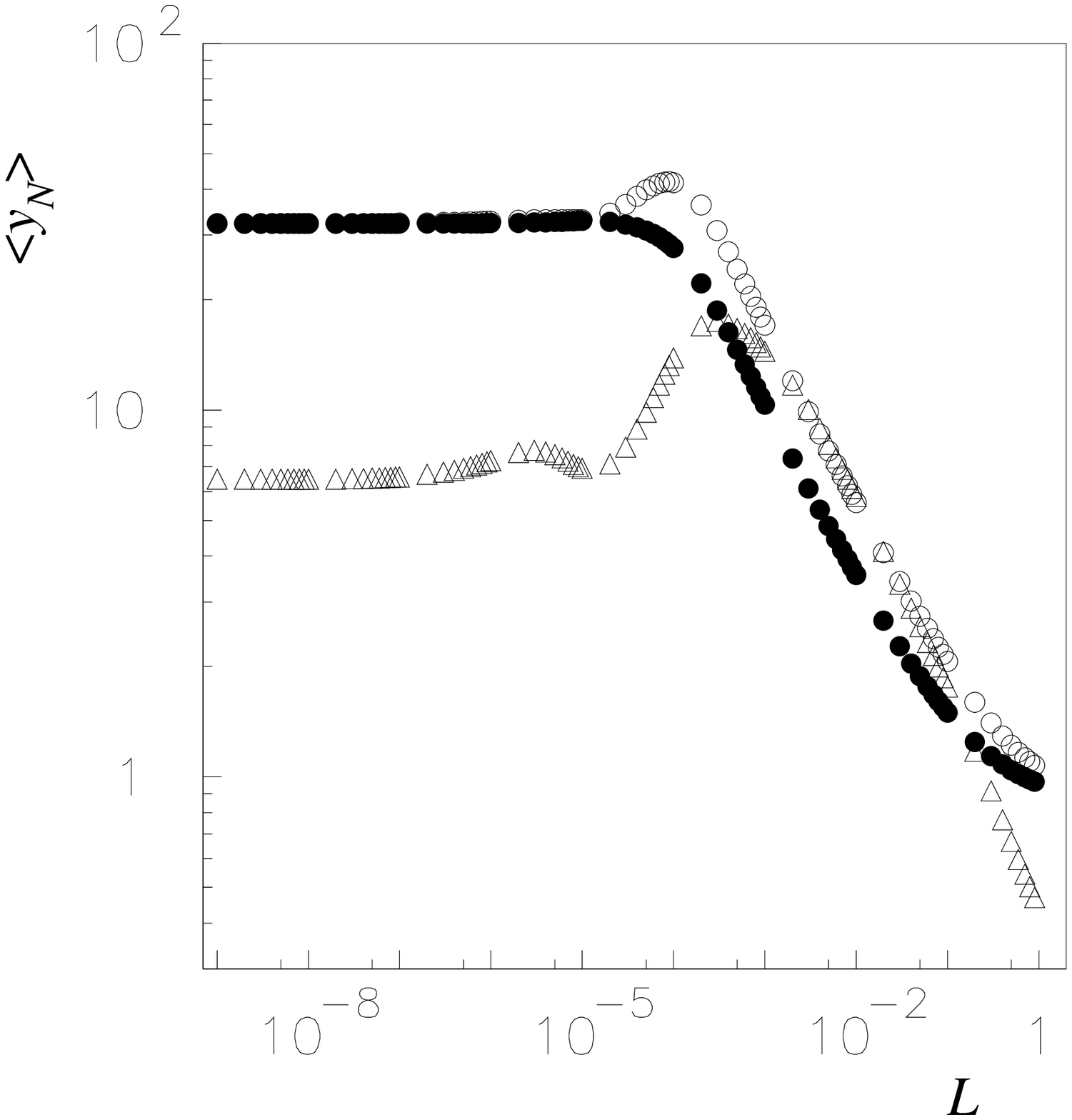}\\*
\end{center}
\caption{Average transverse extension
$\langle y_N\rangle$ as a function of $L$ for $N=100$, $K=10$, and
$\theta_0=\pi/3$ $(\bullet)$, $\theta_0=2 \pi/3$ $(\circ)$ ,
$\theta_0=17 \pi/18$ $(\triangle)$.} 
\label{fig10}
\end{figure}

\begin{figure}[ht]
\begin{center}
   \includegraphics*[width=0.7\textwidth]{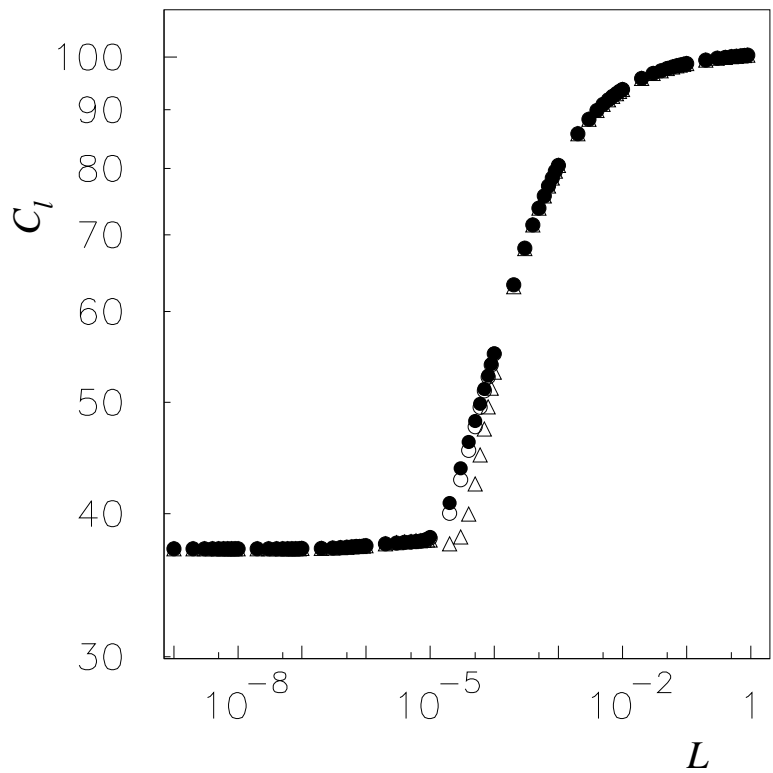}\\*
\end{center}
\caption{Contour length $C_l$ of the average
configuration as a function of $L$ for $N=100$, $K=10$, and
$\theta_0=\pi/180$ $(\bullet)$, $\theta_0=\pi/3$ $(\circ)$,
$\theta_0=2 \pi/3$ $(\triangle)$.} 
\label{fig11}
\end{figure}

\begin{figure}[ht]
\begin{center}
   \includegraphics*[width=0.7\textwidth]{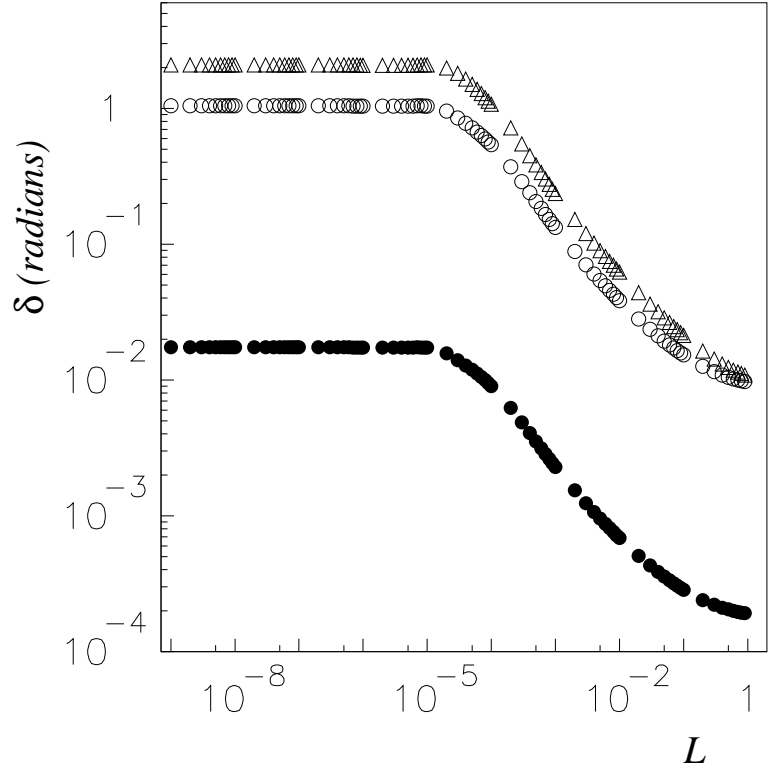}\\*
\end{center}
\caption{Angle $\delta$ between a line
through the end points of the chain and the direction of the force
field as a function of $L$ for $N=100$, $K=10$, and
$\theta_0=\pi/180$ $(\bullet)$, $\theta_0=\pi/3$ $(\circ)$,
$\theta_0=2 \pi/3$ $(\triangle)$.} 
\label{fig12}
\end{figure}

\begin{figure}[ht]
\begin{center}
   \includegraphics*[width=0.7\textwidth]{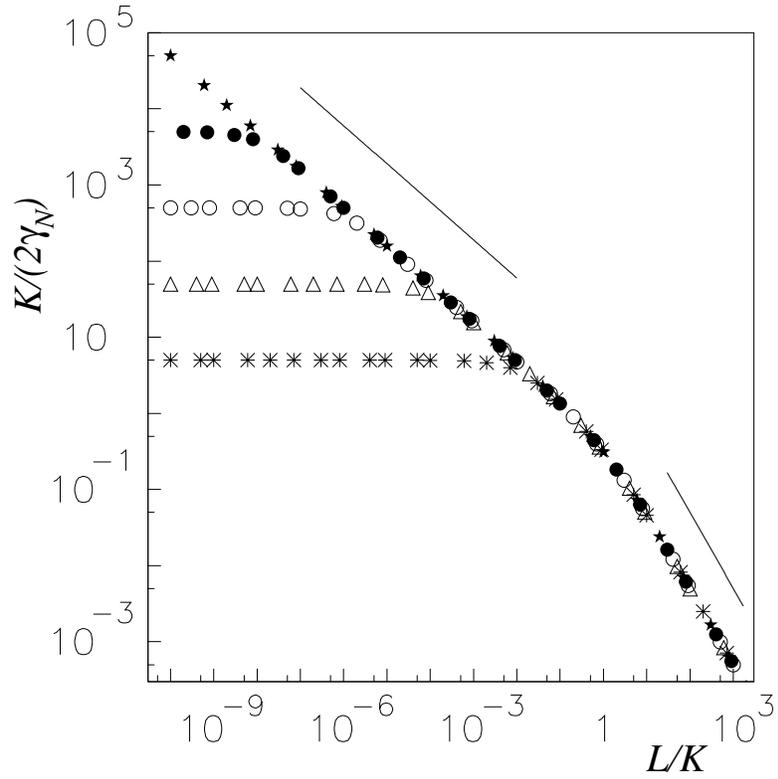}\\*
\end{center}
\caption{$K \langle\theta_N^2\rangle=K/(2\gamma_N)$ for a
polymer pulled at the ends as a function of $L/K$ for $N=10$
($\ast$), $N=100$ ($\triangle$), $N=1000$ ($\circ$), and
$N=10\thinspace 000$ ($\bullet$), together with the exact result
(\ref{fixpoint}) for $N\to\infty$ ($\star$). The straight lines
have slopes $-1/2$ and $-1$, respectively.} 
\label{fig13}
\end{figure}

\begin{figure}[ht]
\begin{center}
   \includegraphics*[width=0.7\textwidth]{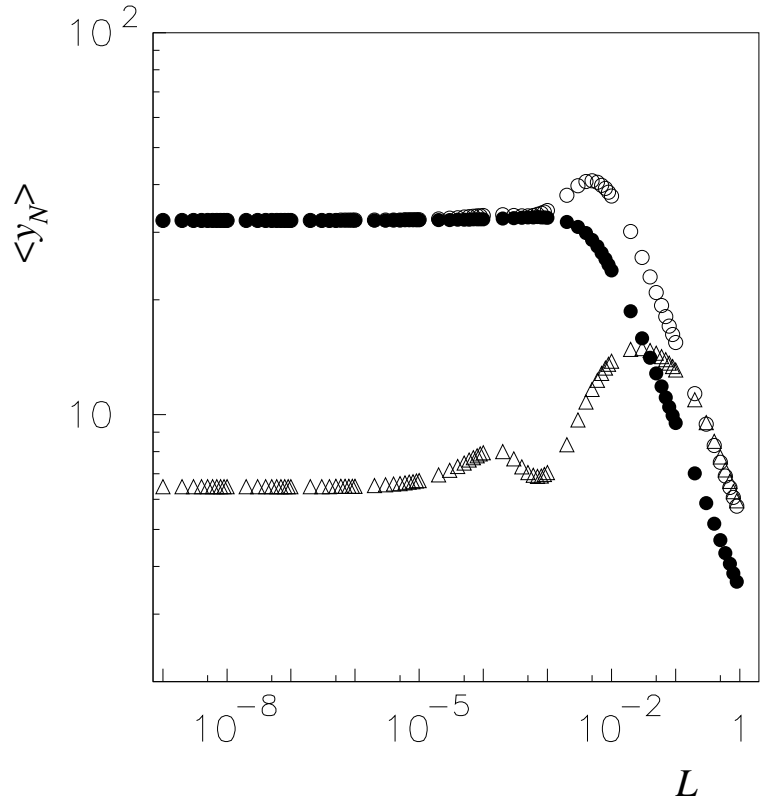}\\*
\end{center}
\caption{Average transverse extension
$\langle y_N\rangle$ of a polymer pulled at its ends as a function
of $L$ for $N=100$, $K=10$, and $\theta_0=\pi/3$ $(\bullet)$,
$\theta_0=2 \pi/3$ $(\circ)$, $\theta_0=17 \pi/18$ $(\triangle)$.} 
\label{fig14}
\end{figure}

\noindent\Large\textbf{Figure Captions} \normalsize \vspace{1.0cm}
\begin{description}
\item{Figure 1}: Average angular fluctuations
$K \langle\theta_i^2\rangle$ for $L/K=0.1$ and $N=100$
$(\bullet)$, $N=500$ $(\star)$, $N=1000$ $(\ast)$.

\item{Figure 2}: Average angular fluctuations
$K \langle\theta_N^2\rangle$ of the final chain segment as a
function of $N$ for $L/K=10^{-5}$ $(\bullet)$, $L/K=10^{-4}$
$(\circ)$, $L/K=10^{-3}$ $(\triangle)$, $L/K=10^{-2}$ $(\star)$,
$L/K=10^{-1}$ $(\ast)$. The straight line has slope 1.

\item{Figure 3}: $N_{\rm min}$ as a function of $L/K$ for
$K=1$ $(\bullet)$, $K=10$ $(\circ)$, $K=100$ $(\triangle)$,
$K=1000$ $(\star)$. The straight line has slope -1/3.

\item{Figure 4}: $K \langle\theta_N^2\rangle=K/(2\gamma_N)$ for a
polymer in a constant force field
as a function of $L/K$ for $N=10$ ($\bullet$), $N=100$ ($\circ$),
and $N=1000$ ($\triangle$), together with the exact result
(\ref{gaminfty}) for $N\to\infty$ ($\star$). The straight lines
have slopes $-1/3$ and $-1$, respectively.

\item{Figure 5}: Average extension,
$\langle x_N \rangle$ of the chain in the direction of the force
field direction as a function of the field strength $L$, for
$K=10$, $N=100$ ($\ast$); $K=10$, $N=1000$ ($\circ$); $K=10$,
$N=10000$ ($\diamond$); $K=100$, $N=1000$ ($\triangle$); $K=100$,
$N=10000$ ($\star$); $K=1000$, $N=10\thinspace 000$ ($\bullet$).
The straight line has slope $-1/2$.

\item{Figure 6}: Average extension $\langle x_N \rangle$ of
the chain in the direction of the force field as a function of the
field strength $L$. The symbols correspond to the same parameters
as in Fig.~\ref{fig5}. The straight line has slope $-1$.

\item{Figure 7}: Average positions
$\langle x_i\rangle,\langle y_i\rangle$ of a chain of length
$N=100$ with $K=1$, $\theta_0=\pi/3$ (lower panel), and
$\theta_0=2 \pi/3$ (upper panel), with $L=10^{-5}$ $(\circ)$,
$L=10^{-4}$ $(\triangle)$, $L=10^{-3}$ $(\star)$, $L=10^{-1}$
$(\bullet)$. The straight line indicates the tilt angle
$\theta_0$.

\item{Figure 8}: Average positions
$\langle x_i\rangle,\langle y_i\rangle$ of a chain of length
$N=100$ with $K=10$, $\theta_0=\pi/3$ (lower panel), and
$\theta_0=2 \pi/3$ (upper panel), with $L=10^{-5}$ $(\circ)$,
$L=10^{-4}$ $(\triangle)$, $L=10^{-3}$ $(\star)$, $L=10^{-1}$
$(\bullet)$. The straight line indicates the tilt angle
$\theta_0$.

\item{Figure 9}: Average transverse extension
$\langle y_N\rangle$ as a function of $\theta_0$ for $N=100$,
$K=10$ and $L=10^{-9}$ $(\ast)$, $L=10^{-5}$ $(\star)$ , and
$L=10^{-3}$ $(\triangle)$ , $L=10^{-2}$ $(\circ)$ , $L=10^{-1}$
$(\bullet)$.

\item{Figure 10}: Average transverse extension
$\langle y_N\rangle$ as a function of $L$ for $N=100$, $K=10$, and
$\theta_0=\pi/3$ $(\bullet)$, $\theta_0=2 \pi/3$ $(\circ)$ ,
$\theta_0=17 \pi/18$ $(\triangle)$.

\item{Figure 11}: Contour length $C_l$ of the average
configuration as a function of $L$ for $N=100$, $K=10$, and
$\theta_0=\pi/180$ $(\bullet)$, $\theta_0=\pi/3$ $(\circ)$,
$\theta_0=2 \pi/3$ $(\triangle)$.

\item{Figure 12}: Angle $\delta$ between a line
through the end points of the chain and the direction of the force
field as a function of $L$ for $N=100$, $K=10$, and
$\theta_0=\pi/180$ $(\bullet)$, $\theta_0=\pi/3$ $(\circ)$,
$\theta_0=2 \pi/3$ $(\triangle)$.

\item{Figure 13}: $K \langle\theta_N^2\rangle=K/(2\gamma_N)$ for a
polymer pulled at the ends as a function of $L/K$ for $N=10$
($\ast$), $N=100$ ($\triangle$), $N=1000$ ($\circ$), and
$N=10\thinspace 000$ ($\bullet$), together with the exact result
(\ref{fixpoint}) for $N\to\infty$ ($\star$). The straight lines
have slopes $-1/2$ and $-1$, respectively.

\item{Figure 14}: Average transverse extension
$\langle y_N\rangle$ of a polymer pulled at its ends as a function
of $L$ for $N=100$, $K=10$, and $\theta_0=\pi/3$ $(\bullet)$,
$\theta_0=2 \pi/3$ $(\circ)$, $\theta_0=17 \pi/18$ $(\triangle)$.

\end{description}
\newpage

\newpage

\newpage

\newpage

\newpage

\newpage

\newpage

\newpage

\newpage

\newpage

\newpage

\newpage

\newpage

\newpage

\end{document}